\documentclass{article}

\usepackage{arxiv}

\usepackage[utf8]{inputenc} % allow utf-8 input
\usepackage[T1]{fontenc}    % use 8-bit T1 fonts
\usepackage{hyperref}       % hyperlinks
\usepackage{url}            % simple URL typesetting
\usepackage{booktabs}       % professional-quality tables
\usepackage{amsfonts}       % blackboard math symbols
\usepackage{nicefrac}       % compact symbols for 1/2, etc.
\usepackage{microtype}      % microtypography
\usepackage{lipsum}
\usepackage{graphicx}
\usepackage{amsmath}
\usepackage{nccmath}
\usepackage[super]{nth}
\graphicspath{ {./images/} }

\title{Macroscopic and Microscopic Characteristics of Networks with Time-variant Functionality for Evaluating Resilience to External Perturbations}

\author{
 Xinyu Gao \\
  Zachry Department of Civil and \\Environmental Engineering\\
  Texas A\&M University\\
  College Station, Texas, 77843 \\
  \texttt{xy.gao@tamu.edu} \\
   \And
 Shangjia Dong \\
  Zachry Department of Civil and \\Environmental Engineering\\
  Texas A\&M University\\
  College Station, Texas, 77843 \\
  \texttt{shangjia.dong@tamu.edu} \\
  \And
 Ali Mostafavi \\
  Zachry Department of Civil and \\Environmental Engineering\\
  Texas A\&M University\\
  College Station, Texas, 77843 \\
  \texttt{amostafavi@civil.tamu.edu} \\
  \And
 Jianxi Gao \\
  Computer Science Department \& \\Network Science and Technology Center\\
  Rensselaer Polytechnic Institute\\
  Troy, New York, 12180 \\
  \texttt{gaoj8@rpi.edu} \\
}

\begin{document}
\maketitle
\begin{abstract}
Knowledge of time-variant functionality of real-world physical, social, and engineered networks is critical to the understanding of the resilience of networks facing external perturbations. The majority of existing studies, however, focus only on the topological properties of networks for resilience assessment, which does not fully capture their dynamical resilience. In this study, we evaluate and quantify network resilience based both on the functionality states of links and on topology. We propose three independent measures---the \textit{failure scaling index} ($FSI$), the \textit{weighted degree scaling index} ($WDSI$), and the \textit{link functionality irregularity index} ($LFII$)---that capture macroscopic, microscopic, and temporal performance characteristics of networks. Accordingly, an integrated \textit{general resilience} ($GR$) metric is used to assess performance loss and recovery speed in networks with time-variant functionality. We test the proposed methods in the study of traffic networks under urban flooding impacts in the context of Harris County, Texas, during Hurricane Harvey using a high-resolution dataset, which contains temporal speed of 20,000 roads every 5 minutes for 5 months. Our results show that link weights and node weighted degrees with perturbed functionality in the traffic network during flooding follow a scale-free distribution. Hence, three proposed measures capture clear resilience curves of the network as well as identify the irregularity of links. Accordingly, network performance measures and the methodology for resilience quantification reveal insights into the extent of network performance loss and recovery speed, suggesting possible improvements in network resilience in the face of external perturbations such as urban flooding.
\end{abstract}

% keywords can be removed

\section{Introduction}
Networks are structures upon which complex behaviors in human, natural, physical, and engineered systems unfold~\cite{albert2000error,cohen2000resilience,gao2016universal}. Due to climate change, the exposure of networks to natural hazards, and even multiple-risk hazards is likely to increase and the damage caused by natural disasters may propagate through underlying networks~\cite{Gallina:2016:review}. A unexpected storm may destroy power transmission towers and lines in power grid network, and the outage caused by it may lead to cascaded failure in transportation networks since the traffic control signals cannot work without power. As society has increasingly higher expectations for the network performance and lower tolerance to performance degradation, to ensure network performance remains at an expected level, it is crucial to guarantee that the network be resilient to external perturbations~\cite{wei2011protecting,zhuang2009earthquake,Dong:2019:Bayesian,Dong:2020,berezin2015localized}. In general, resilience refers to a network’s ability to anticipate, prepare for, and adapt to changing conditions and to withstand, respond to, and recover rapidly from adverse events~\cite{Najjar:1990:network}. 

In particular, networks serve as the skeleton for objects and information to be safely and efficiently transmitted from one end to the other. The study of network structures and properties is essential for examining the resilience of systems under perturbations. Thus, various studies~\cite{Hamedmoghadam:2019:revealing,schafer2018dynamically,Bassola:2019,Gao:2013:percolation} have proposed methods and measures for analyzing network resilience in different research areas. Universal studies about general network resilience, however, focus on synthetic models, e.g. Barabasi-Albert network~\cite{barabasi1999emergence} and Erdos-Renyi network in fictious scenarios~\cite{Duan:2019:universal}. These universal studies and metrics show constrained application in real-life networks, such as epidemic-spreading networks~\cite{Poletto:2013:human}, energy networks~\cite{Robert:2016:analysis}, transportation networks~\cite{LiDaqing:2015:percolation,zeng2019switch,Jiang:2017:spatio,fleurquin:2013:systemic,verma:2014:revealing}, and social networks~\cite{Fan:2020:crowd} due to the strong assumptions in the network topology. With the advancement of technology and the availability of big data, more accurate measures incorporating system dynamics are desired to characterize the network resilience behavior~\cite{gao2016universal}. There is a need to fill the gap between realistic scenarios and analytical models by developing a general analysis method encapsulating both static and dynamic perspectives of network functionality to characterize and quantify the network resilience of various scopes to various perturbations.

A key issue is the specification of measures of performance (MOPs) for network resilience~\cite{Nan:2017:quantitative,ganin:2016:operational}. The assessment of the extent to which MOPs are reduced and the speed at which MOPs recover to a steady-state performance is the standard method for determining network resilience~\cite{Hosseini:2016:review,wan2018resilience}. The majority of the current studies have focused on MOPs related to the topological structure of systems, such as giant connected-component~\cite{Gao:2011:robustness}, network efficiency~\cite{Crucitti:2003,LiQingchun:2019:modeling}, and network centrality (including degree, clustering, and shortest path length)~\cite{Lordan:2014:robustness,Mostafizi:2017:percolation,zhang2020resilience}.

While topological measures provide useful insights regarding network performance in static networks, they ignore the time-variant nature of network functionality whose value is a continuous number. Various engineered networks have components with time-variant functionality (e.g., the flow rate in a power grid network\cite{Davis:2018} and the congestion level in a traffic network~\cite{Chen:2017,Louail:2014:mobile,daqing2014spatial}). For instance, the congested road can still allow vehicles to pass despite low efficiency, while the closed one cannot contribute to the network resilience. These two links, however are considered equivalently as dysfunctional in conventional network structure analysis even though they contribute differently to network resilience.

Therefore, in this study, we examine three network MOPs, including the \textit{failure scaling index} ($FSI$), the \textit{weighted degree scaling index} ($WDSI$), and the \textit{link functionality irregularity index} ($LFII$), that capture important network macroscopic, microscopic, and temporal characteristic. These three MOPs characterize and quantify resilience in such networks under perturbations. The generic nature of the proposed MOPs enables their applications in various contexts in networks with time-variant properties.

\section{Methods}
\subsection{Traffic Network with Time-variant Functionality}
\begin{figure}[!ht]
\centering
\includegraphics[width=\linewidth]{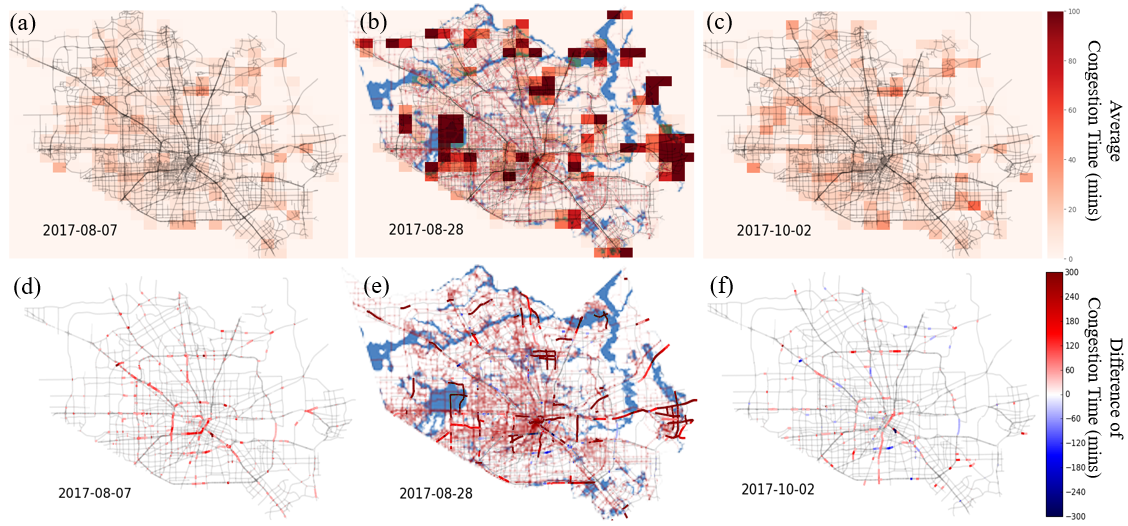}
\caption{\textbf{Maps of average congestion duration as well as the irregularity of congestion duration for the days before, during, and after Hurricane Harvey.} The entire area of Harris County is divided into a 30-by-30 grid (3.3 km-by-2.3 km). The congestion ratio is \textit{q} 0.3. To eliminate the weekly fluctuation, the result of each day is the average of the seven days around the targeted day (i.e. three days before through three days after). The light-blue shaded area indicates the flood area at August 30. (\textbf{a-c}) Average congestion duration of roads in each grid (in minutes). Dark-red shading represents average congestion duration of greater than 100 minutes. (\textbf{d-f}) Difference in the congestion duration compared with the steady-state period (i.e. the average congestion duration for the first week of August). Dark-red shading represents links with more than 300 minutes/day of additional congestion duration during the perturbed state; dark-blue shading represents less congestion by more than 300 minutes/day.}
\label{fig:congetsion}
\end{figure}

Time-variant functionality exists in most real networks, either at the node or link level. In this study, we illustrate the measures and methods in the context of traffic networks exposed to flooding as the external perturbation. Traffic networks are composed of links (road segments) with time-variant functionality (e.g., congestion level). In the context of the traffic network in Harris County during the flooding caused by the 2017 Hurricane Harvey, we plot Figure \ref{fig:congetsion} to show coarse-grained (30-by-30 grid) maps of average congestion duration as well as the irregularity of congestion (i.e., abnormal congestion differentiated from regular fluctuation during the steady state) in the traffic network for three example days before, during, and after the flooding perturbation. The congestion and irregularity identification will be discussed in the method. On August 7, before Harvey, the average congestion duration was distributed evenly with no area experiencing severe congestion and all links performing in the steady-state. On August 28, when Harvey landed, we can see that multiple areas were experiencing severe congestion; a portion of links were more, and some less congested. In the aftermath of Harvey, after a month of recovery of  the traffic network returned to another steady-state. Shading in the figure indicates flooded areas, which are also consistent with traffic congestion. 

The illustration of spatial spread and temporal evolution of functional states in each link demonstrate the fact that the traffic network resilience could not be fully evaluated using static topological measures, and there is a need for measures of performance that capture the time-variant functionality of links. This need motivates the measures and analysis presented in the rest of this paper. 

\subsection{Dysfunction Identification}
We model a real-world system as a weighted network whose links have time-variant functionality. Given a weighted network $G(V, E, W, t)$, it comprises a node set $V = \{ v_1,\ldots ,v_n \} $ of size $N$ and a link set $E = \{ e_{ij} (t),i,j=1,\ldots ,n \} $, where $e_{ij}(t)$ represents the link between node \textit{i} and node \textit{j} at time \textit{t}. $W(t)$ is an $n\times n$ symmetric matrix, where $w_{ij}(t)$ represents the weight of the link $e_{ij}$ at time $t$. A link represents a system component and the weight captures the functionality of each link at different time. For example, the level of power flow in transmission lines varies at hourly as the demand or the congestion level in traffic networks fluctuate during rush hours and non-rush hours. Such varying network performance will be exacerbated in the face of perturbations caused by natural and human-made disasters. To characterize the functional failure of the links, we define $F_{oij}$ and $F_{ij}(t)$ to represent the optimal functionality (e.g., the maximum average speed on traffic networks without any congestion) and the actual functionality of link $e_{ij}$ at time $t$, respectively. Then the ratio of the actual functionality to optimal functionality of the links is defined as the extent to which a link’s functionality is disrupted. Accordingly, we define a dysfunction threshold $q$ (which is congestion ratio in the traffic network), below which a link $e_{ij}$ is considered as failed. We can formulate the dysfunction of links by Eq. (\ref{eq:1}).

\begin{equation}
\label{eq:1}
Dysfunction(i,j,t)=
\begin{cases}
0  &F_{ij}(t)/F_{oij}>q\\
1  &F_{ij}(t)/F_{oij}<q
\end{cases}
\end{equation}

This threshold enables modeling dysfunction based on changes in the time-variant functionality of links. For example, in traffic networks, if a road experience congestion and the function of the road, $F_{ij}(t)$, drops below a threshold during rush hours, the link is considered to be dysfunctional. Hence, the links may fail for two reasons:(i) external perturbations and (ii) the network's internal dynamics induced dysfunctionality. To characterize the effects of the temporal functional failures in links, we introduce the time variable $DP_{ij}(t_m,t_n)$ to measure the dysfunction proportion of a link $e_{ij}$ where $t_m$ and $t_n$ are the start time and end time, respectively. $DP_{ij}(t_m,t_n)$ enables capturing the link functionality during a specific time period and can be formulated as Eq. (\ref{eq:2}):

\begin{equation}
\label{eq:2}
DP_{ij}(t_m,t_n)=\frac{\sum_{t_m}^{t_n}Dysfunction(i,j,t)}{t_n-t_m}
\end{equation}

This variable captures the proportion of time that a link’s functionality drops to a level which is considered to be failed. In the proposed methodology, $DP_{ij}(t_m,t_n)$ is used as the weight of links during that period. For example, a period of 60 minutes from 8 a.m. to 9 a.m. is selected, if a link’s functionality falls below the thresholds over 45 minutes, the weight of the link would be 0.75.

\subsection{Macroscopic Network Performance Characterization}
Using the series data of $DP_{ij}(t_m,t_n)$, we can analyze the data from both the macroscopic and microscopic ways. From a macroscopic perspective, the network is considered as a whole with less focus on the individual component. Performance information can be aggregated into one index using both statistics and network theory, regardless of other details on every single link, e.g., longitude, latitude, and length. From the microscopic perspective, each component will be analyzed separately to capture the network performance at the component level. The abnormal links can be found and identified, geographic information of which can be used to map vulnerable areas and components in future analysis. These two approaches complement each other by highlighting the different properties of network performance.

\subsubsection{\textit{Failure Scaling Index} (\textit{FSI})}
The first macroscopic characteristic is the \textit{FSI}. After assigning the $DP_{ij}(t_m,t_n)$ as the weight of each link, we can plot the distribution of link weights for the entire network in log-log scale to examine the overall network performance. Jiang et al.\cite{Jiang:2017:spatio} found that the link dysfunction distributions in traffic networks follow a power-law distribution, so we hypothesize that the link weight distribution in networks whose links have time-variant functionality follows the relationship in Eq. (\ref{eq:3}): 

\begin{equation}
\label{eq:3}
p(t_m,t_n) \sim w(t_m,t_n)^{-\alpha}
\end{equation}

where $-\alpha$ is the exponent, $p$ is the probability density of the link weight, and $w$ is the weights of links. Accordingly, we reduce the scales in both  using logarithmic scales in both dimensions; thus the relationship is converted into $log p(t_m,t_n) \sim {-\alpha}log w(t_m,t_n) $. Here, we define the inverse of the slope $\alpha$ as the \textit{FSI}. By plotting distribution of the link weights in logarithmic space and fitting the distribution with a linear relationship, we can derive the \textit{FSI} using the ordinary least square method. If the hypothesis of power-law is not suitable in some real-world network cases, other statistical parameters representing the shape and position of the distribution of link weights can also be examined, a methods also applicable to \textit{WDSI}.

\subsubsection{\textit{Weighted degree scaling index} (\textit{WDSI}) }
The second macroscopic characteristic is the \textit{WDSI}. In fact, a link functionality state also affects the state of the node to which it is connected. Aggregating link weights connected to a node, a weighted degree can be derived for all nodes in a network, as shown in Eq. (\ref{eq:4}):

\begin{equation}
\label{eq:4}
d_k(t_m,t_n)=\sum_{i\subseteq N(k)} w_{ik}(t_m,t_n)
\end{equation}

where, $w_{ik}$ is the weight of the links connecting to nodes $v_i$ and $v_k$, $N(k)$ is a set of nodes that connect to node $v_k$. Similarly, the probability distribution ($p'(t_m,t_n)$) of the weighted degree of nodes in a network can also be determined using Eq. (\ref{eq:5}):
\begin{equation}
\label{eq:5}
p'(t_m,t_n) \sim d_k(t_m,t_n)^{-\beta}
\end{equation}

where $-\beta$ is the exponent, and $d_k$ is the weighted degree of nodes. Accordingly, we map the function on a logarithmic scale. Similar to $FSI$, we hypothesize the weighted degree distribution follows a power-law pattern, and thus the inverse number of the slope, $\beta$ is defined as the $WDSI$.

\subsection{Microscopic Network Performance Characterization}

\subsubsection{\textit{Link Functionality Irregularity Index} (\textit{LFII}) }
In addition to the macroscopic measures of performance, we also define a microscopic measure of network performance, \textit{LFII}. Under steady state, a link’s functionality (i.e., $DP_{ij}(t_m,t_n)$) fluctuates due to the inherent dynamics of the network. Such fluctuation usually falls within an interval defined by an upper-bound in Eq. (\ref{eq:6}) and a lower-bound threshold in Eq. (\ref{eq:7}): 

\begin{equation}
\label{eq:6}
\psi_{Uij} = min\{ DP_{oij}+max\{ \lambda \overline{DP_o},(\lambda -1)DP_{oij} \}, 1 \}
\end{equation}

\begin{equation}
\label{eq:7}
\psi_{Lij} = max\{ DP_{oij}-max\{ \lambda \overline{DP_o},\frac{\lambda -1}{\lambda}DP_{oij} \}, 0 \}
\end{equation}

where $\psi_{Uij}$ and $\psi_{Dij}$ is the upper- and lower- bound threshold for the functionality steady-state of link $i-j$, respectively. $DP_{oij}$ indicates the original status of link $i-j$ and $\overline{DP_o}$ is the average of $DP_{oij}$ for all links. Tolerance index $\lambda$ is a positive variable that can be used to determine the interval between two thresholds, which is suggested to vary in the range of [1,3]. As $\lambda$ increases, the steady-state interval increases as well. Usually the first week of investigation period can be used as the start time for calculating $DP_{oij}$, which can minimize the weekly fluctuation of link functionality. However, under external perturbations, the functionality of links would fall outside of the steady-state fluctuation intervals. This irregularity indicate a link’s incapacity in coping with external perturbations (because of either functional of physical failures). Hence, we define \textit{LFII}-the proportion of links whose performance fall within the steady-state thresholds-as the measure for the network to capture the extent to how many links (subjected to external perturbations) are behaving similarly to their regular steady-state performances. \textit{LFII} can be calculated using Eq. (\ref{eq:8}) and Eq. (\ref{eq:9}):

\begin{equation}
\label{eq:8}
LFII(t_m,t_n)=\frac{\sum_{\forall i\neq j} \phi (DP_{ij} (t_m,t_n))}{\sum_{\forall i \neq j} e_{ij}}
\end{equation}

\begin{equation}
\label{eq:9}
\phi (DP_{ij} (t_m,t_n)) =
\begin{cases}
1  &\psi_{Lij}\leq DP_{ij} (t_m,t_n) \leq \psi_{Uij} \\
0  &Otherwise
\end{cases}
\end{equation}

where $DP_{ij}(t_m,t_n)$ is dysfunction proportion defined before, and $\phi$ is a function determining whether the link behaviors is similar to its regular performance by comparing it with $\psi_{Uij}$ and $\psi_{Dij}$. Since $LFII$ is defined as a proportion of links, the value of \textit{LFII} fall into the range from 0 to 1.

\subsection{Characterizing Network Resilience}
\begin{figure}[!ht]
\centering
\includegraphics[width=0.5\linewidth]{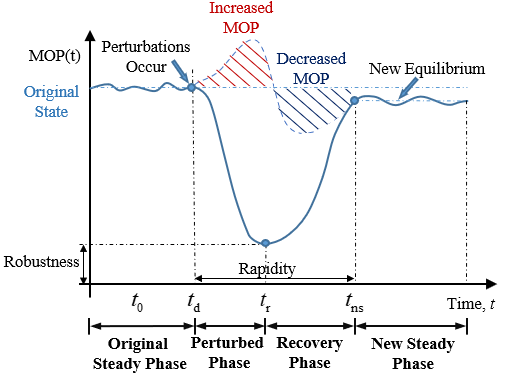}
\caption{System resilience transitions and phases.}
\label{fig:x-resilience curve}
\end{figure}

The \textit{FSI}, \textit{WDSI}, and \textit{LFII} are three fundamental MOPs for characterizing and quantifying network resilience to perturbations. As suggested by other studies related to network resilience~\cite{Nan:2017:quantitative,Hosseini:2016:review,wan2018resilience,Boccaletti:2006}, we examine \textit{robustness} (\textit{R}), \textit{rapidity}) ($RAPI_{PP}$) for perturbed phase and $RAPI_{RP}$ for recovery phase), \textit{average time of performance loss} (\textit{ATPL}), and \textit{recovery ability} (\textit{RA}). We also adopt an integrated measure, \textit{general resilience} ($GR$), proposed by Nan and Sansavini~\cite{Nan:2017:quantitative}. To elucidate these metrics, we examine four different stages of perturbation as shown in the resilience curve in Figure ~\ref{fig:x-resilience curve}, including the initial regular phase, the disruption phase, the recovery phase, and the new steady phase, which are divided by three critical time points $t_d$, $t_r$, and $t_{ns}$, where $t_d$ represents the time when perturbations occur, $t_r$ indicates the time when MOP reaches its minimum value during perturbations, and $t_{ns}$ denotes the time when the MOP recovers from perturbations and achieves another steady status. Accordingly, we consider the increased $MOP$ and decreased $MOP$ as both non-resilient behaviors, because a temporary increase in network $MOPs$ also indicates a non-steady state. As shown in Figure \ref{fig:x-resilience curve}, the blue dashed line represents a situation when perturbation occurs. The MOP increases at first, then decreases and recovers to a new steady phase. 

These six metrics describe different aspects of network resilience when perturbations occur. In order to make each metric comparable, $MOP$ need to be normalized and the original state is set as 1. $R$ is defined as the minimum value of MOP during perturbation period as in Eq. (\ref{eq:10}):
\begin{equation}
\label{eq:10}
R = min\{MOP(t)\}(\mbox{for  } t_d \leq t \leq t_{ns})
\end{equation}

\textit{RAPI} is introduced to capture how fast the networks would respond to perturbations. To be specific, the rapidity in perturbed phase $RAPI_{PP}$ is determined using Eq. (\ref{eq:11}) to represent how fast $MOP$ drops. Similarly, the rapidity in recovery phase $RAPI_{RP}$ is determined using Eq. (\ref{eq:12}) to represent how fast $MOP$ recovers.

\begin{equation}
\label{eq:11}
RAPI_{PP} = \frac{MOP(t_d)-MOP(t_r)}{t_r-t_d}
\end{equation}

\begin{equation}
\label{eq:12}
RAPI_{RP} = \frac{MOP(t_{ns})-MOP(t_r)}{t_{ns}-t_r}
\end{equation}

\textit{ATPL} is defined as the total \textit{performance loss} (\textit{PL}) divided by perturbed duration ($t_{ns}-t_d$). In most studies, \textit{PL} is calculated using $\int_{t_d}^{t_{ns}} MOP(t_0)-MOP(t) dt$, and this equation does not take the increased MOP into account. As the blue dashed line shown in Figure \ref{fig:x-resilience curve}, we use the summation of both red and blue shadow areas as the performance loss instead of simply deducting them. Therefore, the absolute value of the deduction between $MOP(t_0)$ and $MOP(t)$ are used when calculating \textit{ATPL} as shown in Eq. (\ref{eq:13}). 

\begin{equation}
\label{eq:13}
ATPL = \frac{\int_{t_d}^{t_{ns}} |MOP(t_0)-MOP(t)| dt}{t_{ns}-t_d}
\end{equation}

\textit{Recovery ability} ($RA$) in Eq. (\ref{eq:14}) is defined to quantify how much the $MOP$ recover from the minimum value.
\begin{equation}
\label{eq:14}
RA = |\frac{MOP(t_{ns})-MOP(t_r)}{MOP(t_0)-MOP(t_r)}|
\end{equation}

Based on aforementioned equations and definitions, each metric has either positive ($R$, $RAPI_{RP}$, and $RA$) or negative effect ($RAPI_{PP}$, $ATPL$) on network resilience. Another integrated metric, $GR$, is used to calculate a no-weighting factor with no introduced bias with Eq. (\ref{eq:15}). Since all the metrics used to calculate $GR$ are all non-negative metrics, $GR$ is a non-negative metric whose value equals zero when:

\begin{itemize}
\item MOP drops to zero after perturbations ($R$=0)
\item MOP reaches its lowest level immediately when perturbations occur ($RAPI_{PP} \rightarrow \infty$)
\item MOP doesn’t recover at all after perturbations ($RAPI_{RP} = 0$)
\end{itemize}

\begin{equation}
\label{eq:15}
GR = f(R,RAPI_{PP},RAPI_{RP},ATPL,RA)=R\times \frac{RAPI_{RP}}{RAPI_{PP}} \times ATPL^{-1} \times RA
\end{equation}

\section{Results}

The proposed measures and method are examined in a study of traffic network resilience under urban flooding in Harris County, Texas, during Hurricane Harvey in 2017, which stalled over Harris County August 26 through 28, and is one of the costliest natural disasters in United States history~\cite{Van:2017:HarveyRainfall}. A high-resolution dataset (i.e., the average speed on road segments at 5-minute intervals during the 92 days from August 1, 2017, to October 31, 2017, was obtained and used in conjunction with the regional road network topology data \cite{Txdot2020} to build the traffic network model, which contains 15,390 nodes and 19,712 links. Each link contains road dynamics, including average travel speed; reference speed; road closure status for every 5-minute-interval; and road static information, such as road name, coordinates of start and end intersections, and link length. Reference speed is equivalently used as the free-flow speed, which is the speed driven on a road when there is no congestion. In this case, we consider a road to have optimal travel functionality when the average speed on the link is equal to the reference speed. The reference speed is assigned as $F_{oij}$ ,and the current speed as $F_{ij}$. 

\subsection{Macroscopic Characteristics of a Traffic Network: \textbf{\emph{FSI}} and \textbf{\emph{WDSI}}} 
\begin{figure}[!ht]
\centering
\includegraphics[width=\linewidth]{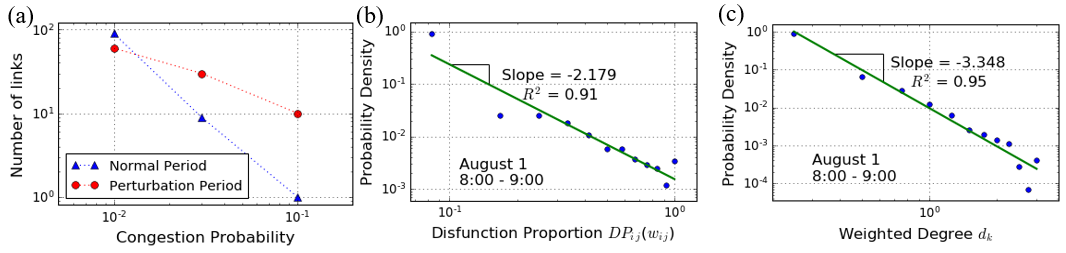}
\caption{\textbf{Calculation of \textbf{\emph{FSI}} and \textbf{\emph{WDSI}}.} \textbf{(a)} The link functionality distribution follows a scale-free distribution and the slope of the curve can represent the performance of network. The blue line indicates the network in a normal period; the red line indicates the network in a perturbation period. \textbf{(b)–(c)} The probability distributions of link weights (functionality) $w_{ij}$ and the nodes weighted degree $d_k$ on August 1 from 8:00 a.m. to 9:00 a.m. The green lines indicate the fitted line to the scattered data using least square method. Slopes and R-square values are shown in each subplot.}
\label{fig:2-demo}
\end{figure}

Macroscopic characteristics are defined as the inverse of the slope of link dysfunction distributions in logarithmic space. For example, in a traffic network with 100 links, during a normal period, 90 links will have a congestion probability of 1\%, nine links with 3\%, and one link with 1\%. The logarithmic relationship of congestion probability and link number is plotted in Figure \ref{fig:2-demo}(a) using a blue dotted line. When an external perturbation occurs, the distribution of link congestion probability changes to 60 links with 1\%, 30 links with 3\%, and ten links with 1\%, respectively. The relationship is also plotted in Figure \ref{fig:2-demo}(a) using a red dotted line. Some links become more congested in the perturbation period, the downward slope of which is less pronounced.  (A shallower downward slope signifies a smaller MOP, since the inverse of the slope is used.) Here, we present an example of calculating the $FSI$ and $WDSI$ on August 1 from 8:00 a.m. to 9:00 a.m. with congestion ratio \textit{q} = 0.3. For each link, $DP_{ij}$ (the congestion probability) is calculated, which is also used as the link weight $w_{ij}$. The weighted degree of each node $d_k$ is then derived. As shown in Figure \ref{fig:2-demo}(b) and (c), the probability distributions of $w_ij$ and $d_k$ follow the power-law as they show a linear relationship in logarithmic space. In this case, $FSI$ equals to 2.179 and $WDSI$ equals to 3.348, which are the inverse of the slopes of two fitted lines.

\begin{figure}[!ht]
\centering
\includegraphics[width=\linewidth]{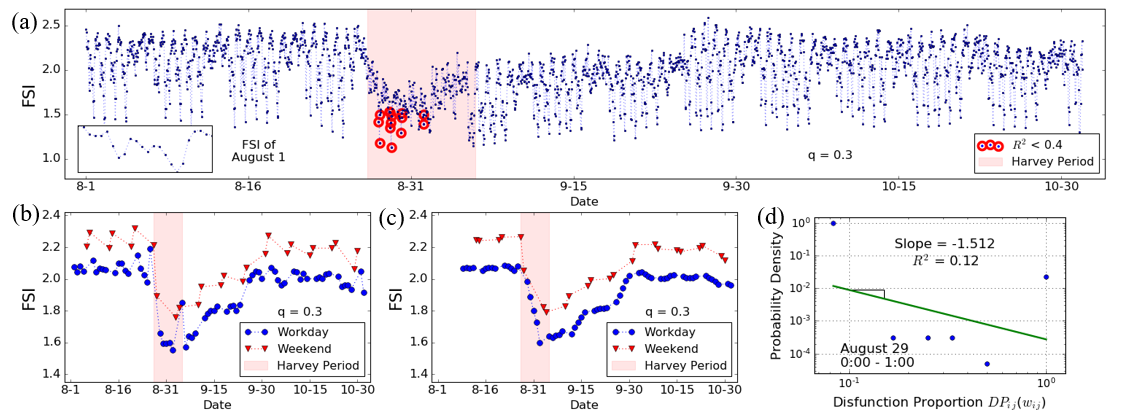}
\caption{\textbf{Change of \textit{FSI} for Harris County traffic network before, during and after Hurricane Harvey.} The congestion ratio $q$ is 0.3. The light-red shaded area indicates the period of Hurricane Harvey (August 26 through September 4). \textbf{(a)} Each point indicates an hourly $FSI$, and all points are linked by a dashed line to show the changing pattern. Red circles indicate that this $FSI$ has an R-squared value less than 0.4 in the linear regression process. The subplot shows the $FSI$ change in August 1st. \textbf{(b)} The daily average $FSI$ separated by workday and weekend. \textbf{(c)} The weekly moving average $FSI$ separated by workday and weekend. \textbf{(d)} Sample of $FSI$ with R-squared value less than 0.4.}
\label{fig:3-Hourly FSI}
\end{figure}

\begin{figure}[!ht]
\centering
\includegraphics[width=\linewidth]{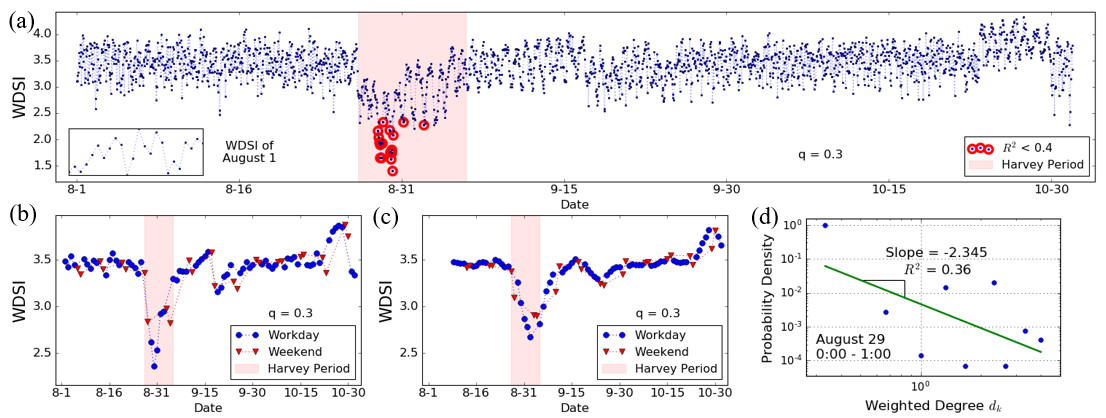}
\caption{\textbf{Change of \textit{WDSI} for Harris County traffic network before, during and after Hurricane Harvey.} The congestion ratio $q$ is 0.3. The light-red shaded area indicates the period of Hurricane Harvey (August 26 through September 4). \textbf{(a)} Each node indicates an hourly $WDSI$, and all nodes are linked by dashed line to show the changing pattern. The data with red circle indicates that this $WDSI$ has an R-squared value less than 0.4 in the linear regression process. The subplot shows the $WDSI$ change in August 1st. \textbf{(b)} Daily average $WDSI$ separated by workday and weekend. \textbf{(c)} Weekly moving average $WDSI$ separated by workday and weekend. \textbf{(d)} Sample of $WDSI$ with R-squared value less than 0.4.}
\label{fig:4-Hourly WDSI}
\end{figure}

Repeating the previous procedure, we calculated hourly $FSI$ and $WDSI$ when $q$ is equal to 0.3, as shown in Figure \ref{fig:3-Hourly FSI}(a) and Figure \ref{fig:4-Hourly WDSI}(a). The normalization was not conducted for MOPs, since it is reasonable to analyze the sensitivity of each MOP to parameters using the original value. Since we use linear regression calculation to determine $FSI$ and $WDSI$, the R-squared values reflect the presence of a power-law scaling in link weight and node-weighted degree distributions. Two examples with poor R-squared values are plotted in Figure \ref{fig:3-Hourly FSI}(d) and Figure \ref{fig:4-Hourly WDSI}(d). The period of Hurricane Harvey is defined from August 26 to September 4 (10 days). The hurricane landed in Harris County on August 26. Floodwaters receded almost a week after the heavy precipitation. The Harvey period is highlighted using a light-red shaded area in all the figures.

Both $FSI$ and $WDSI$ remained stable during the steady state before Harvey. During the steady-state period, $FSI$ values vary within the range of [1.5, 2.5], and $WDSI$ values vary within the range of [2.5, 4.0]. During the perturbation stage due to Harvey, both $FSI$ and $WDSI$ drop significantly then recover from the perturbation and reach another stable status, lower than the pre-Harvey steady state. We also calculated average values for every 24-hour period and determined the daily average value of $FSI$ and $WDSI$ to eliminate noise (see Figure \ref{fig:3-Hourly FSI}(b) and Figure \ref{fig:4-Hourly WDSI}(b)). We separated and compared workday and weekend $FSI$ and $WDSI$. The daily patterns for $FSI$ show more divergence between workday and weekend than the $WDSI$ regular patterns. In the first seven days of August (during the steady-state), the average value of $FSI$ for the weekends is 2.25,  8.7\% higher than the average values for the workdays (2.07). For the same period, however, the average weekend $WDSI$ value is 3.38, similar to the average value of workdays (3.48). In Figure \ref{fig:3-Hourly FSI}-(b) and Figure \ref{fig:4-Hourly WDSI}-(b), the daily average values for $FSI$ and $WDSI$ fluctuate over the course of weeks. We calculated the weekly average values of $FSI$ and $WDSI$ (Figure \ref{fig:3-Hourly FSI}-(c) and Figure \ref{fig:4-Hourly WDSI}-(c)) 
to eliminate the weekly fluctuation. Before Harvey, workday $FSI$ is around 2.07. When Harvey occurred, $FSI$ started dropping on August 28 and reached its minimum value of 1.60 on September 1 (a decrease of 23\%). Then after Harvey, it recovers and reaches 1.81 (an increase of 13\%) on September 15 and remains for a week. Starting from September 25, workday $FSI$ recovers again and reaches 2.0 (an increase of 10\%) after October 2. Weekend $FSI$ shows a similar pattern but has a higher value than a workday. As for $WDSI$, the difference between workday and weekend is not significant. Before Harvey, the workday $WDSI$ value is around 3.45. Hurricane Harvey starting  dropping from August 26 and reached its minimum value of 2.68 on September 1 (a decrease of 22\%). Then after Harvey, it recovers and reaches 3.43 (an increase of 28\%) on September 10, then remains at that value for a week. Starting from September 15, it drops again to 3.23 (a decrease of 6\%) on September 24, then recovers to 3.48 (an increase of 8\%) on October 1, close to the value during the pre-Harvey steady state period. The weekly averaged results show clear resilience curves that can be used to quantify network resilience in the next step.

\begin{figure}[!ht]
\centering
\includegraphics[width=\linewidth]{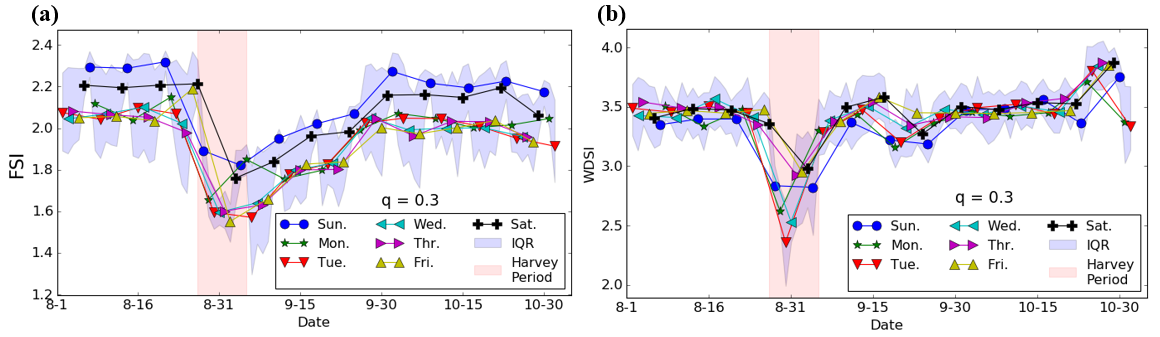}
\caption{\textbf{Daily change of averaged \textbf{\emph{FSI}} and \textbf{\emph{WDSI}} grouped by day of week.} The congestion ratio $q$ is 0.3. The light-red shaded area indicates the period of Hurricane Harvey (August 26 through September 4). The light blue shaded area indicates the inter-quartile range of daily index series. \textbf{(a)} $FSI$ change by day of week. \textbf{(b)} $WDSI$ change by day of week.}
\label{fig:5-Day of week}
\end{figure}

In addition to the moving average calculation, we also examined each day of the week separately and show the \nth{25} and \nth{75} percentile (inter-quartile range, i.e., IQR) which is indicated by a light-blue shaded area of each day for $FSI$ and $WDSI$ value over 24 hours in Figure \ref{fig:5-Day of week}. By comparing these two subplots, the difference between workday and weekend $FSI$ becomes noticeable, which differs from $WDSI$. The IQR of $WDSI$ is smaller than that of $FSI$, indicating that $WDSI$ is more stable than $FSI$ in daily network performance measurement since $FSI$ captures daily traffic dynamics (i.e., fluctuation in the congestion levels of links) more than $WDSI$. According to the definitions, $FSI$ and $WDSI$ describe the distribution of link weight and nodes' weighted degree. Link weight describes link functionality, while weighted degree encapsulates both link functionality and network topology. The link functionality of workdays differs from that of weekends, since people’s destinations shift from the workplace to other places. As the total traffic volume drops during the weekend, $FSI$ changes accordingly. As for $WDSI$, such difference in link functionality is redistributed by the network topology. Therefore $WDSI$ doesn’t show a significant difference between workday and weekend. 

\begin{figure}[!ht]
\centering
\includegraphics[width=0.8\linewidth]{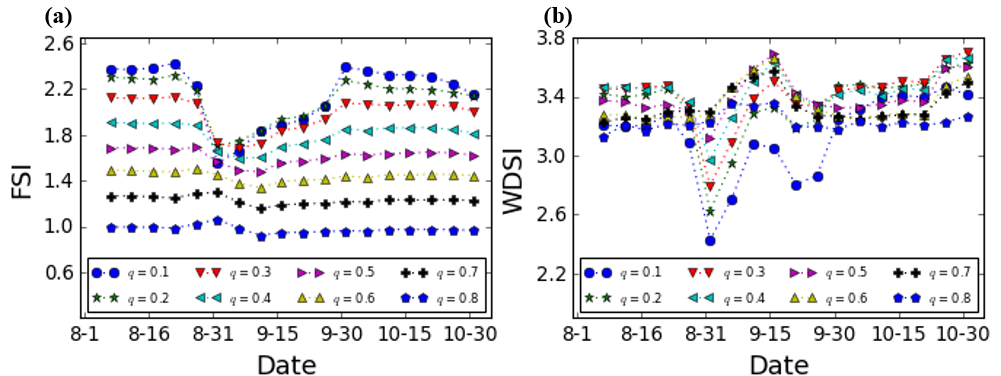}
\caption{\textbf{Daily change of averaged \textbf{\emph{FSI}} and \textbf{\emph{WDSI}} for Harris County traffic network before, during and after Hurricane Harvey, using different congestion ratio \textbf{\emph{q}}.} \textbf{(a)} $FSI$ change by date. \textbf{(b)} $WDSI$ change by date.}
\label{fig:6-different q}
\end{figure}

Varying congestion ratio $q$ would lead to different $FSI$ and $WDSI$ since $q$ determines the identification of functionally failed links. To examine the sensitivity of $FSI$ and $WDSI$ to $q$, we calculated weekly moving average values of $FSI$ and $WDSI$ as shown in Figure \ref{fig:6-different q}. In the sensitivity analysis, $q$ changes from 0.1 to 0.8 with increment of 0.1. The results show that $FSI$ is sensitivity to changes in the $q$ values during steady-state, while $WDSI$ shows sensitivity to variation in the $q$ values during Harvey perturbation period. As shown in Figure \ref{fig:6-different q}, $FSI$ and $WDSI$ have similar curve pattern when $q$ is small. When $q$ increases, $FSI$ drops, and decreases quicker in normal time than in perturbation time; however, $WDSI$ raises in perturbation time and fluctuates slightly in normal time. The congestion ratio $q$ is the dysfunction threshold. When $q$ is equal to 0.1, it means a link would be denoted as dysfunctional when its current functionality is lower than 10\% of its optimal functionality. When $q$ increases, more dysfunctions will be captured cumulatively for links, which means $DP_{ij}$ increases accordingly from 0 to 1 for each connection. If $q$ is more than 1.0, all links will be considered as dysfunctional at all time points, which means that each $DP_{ij}$ becomes 1 (the situation in which the current speed is higher than reference speed is not considered here). In this case, the network would be a static network with all weights in links equal to 1. Therefore, a proper value of $q$ is critical for the resilience analysis, and from the result, we suggest the $q$ in range of [0.1,0.5], beyond which the dysfunctionality will be hard to capture, and the resilience curves will be unclear.

\subsection{Microscopic Characteristics of Traffic Network: \textbf{\emph{LFII}}}

In addition to $FSI$ and $WDSI$, MOPs capture macroscopic characteristics of the traffic network. We examined $LFII$ as a microscopic MOP. Unlike the previous two macroscopic features focusing on the whole network’s performance, \textit{LFII} focuses on performance of every single link. In a steady state, the system will act in its own patterns, such as daily rush hours and more congestion during the work week than on weekends. In the face of perturbations, however, some links would show irregularities due to the disturbance. For example, some links would have more traffic due to failures of other connections or due to  travel behavior change. Therefore, the $LFII$ is examined to assess the microscopic performance of links and to reveal the spatial distribution of failures (i.e., road closures or increased congestion) in facing flooding perturbations (Figure \ref{fig:congetsion}). 

\begin{figure}[!ht]\centering
\includegraphics[width=0.85\linewidth]{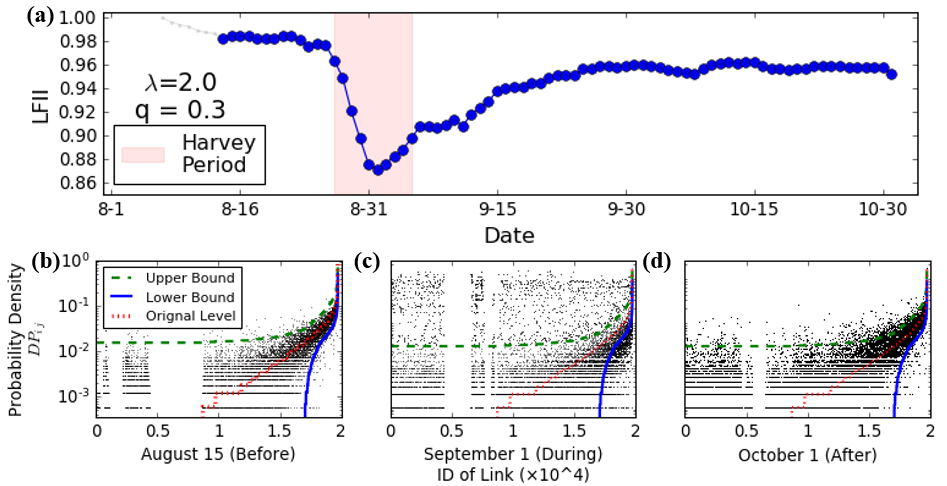}
\caption{\textbf{Daily change of \textit{link functionality irregularity index} (\textbf{\emph{LFII}}) for Harris County traffic network before, during, and after Hurricane Harvey.} The tolerance index $\lambda$ is equal to 2.0. The congestion ratio $q$ is equal to 0.3. The light-red shaded area indicates the period of Hurricane Harvey (August 26 through September 4).\textbf{(a)} $LFII$ change by date. \textbf{(b–(d)} Scatter plots of $DP_{ij}(t_n)$ on August 15, September 1, and October 1, which are before, during and after Harvey respectively. Green line and blue line are upper- and lower-boundaries, respectively, determined based on $\lambda$. Red line indicates the baseline$DP_{oij}$. During the steady state (August 15), the functionality fluctuation of most links falls within the lower and upper bound intervals. The flood-induced perturbations (September 1) cause irregularities in link functionality to fall outside the boundaries. As the network recovers (October 1), the functional states of links fall back almost  within the fluctuation boundaries.}

\label{fig:7-LFII}
\end{figure}

To eliminate weekly variations in the functionality of each link, we used the first week to calculate the original state of a link i-j, $DP_{oij}$. Then $LFII$ was calculated daily starting from August 8, with the dysfunction proportion of each day being the rolling  average of 7 days around the targeted day (i.e., three days prior and three days following). As shown in Figure \ref{fig:7-LFII}(a) ($q = 0.3$ and $\lambda = 2$ are used). $LFII$ value starts dropping from 1.00 on August 7, and reaches its stable state with a value of 0.982 starting August 15. This drop is caused by the inherent fluctuation of the link's functionality. When hurricane perturbation happens, $LFII$ decreases significantly from August 26 to a minimum value of 0.871 on September 1, a decrease of 11\%. Then, $LFII$ recovers from the perturbation and reaches another stable state after October 1 with a value of 0.958, an increase of 10\%. The new steady-state has a $LFII$ value that is 2.4\% lower than the pre-Harvey steady-state, resulting from  changes in mobility demand/patterns and the temporary closure of businesses or flooding of residential areas. We use three days as an example, August 15, September 1 and October 1, which represent states before, during, and after Hurricane Harvey, respectively, to demonstrate the change of $DP_{ij}(t)$ as shown in Figure \ref{fig:7-LFII}(b)-(d). The red, green, and blue lines indicate the $DP_{oij}$, upper- and lower-bound thresholds with $\lambda$ equal to 2.0. Each number on the x-axis indicates a unique link in the network, and the corresponding y-value indicates the $DP_{ij}(t)$ of that link. The $LFII$ will be the proportion of points that fall inside the upper- and lower-bound thresholds, which means these links perform in their regular manner.

\begin{figure}[!ht]
\centering
\includegraphics[width=0.75\linewidth]{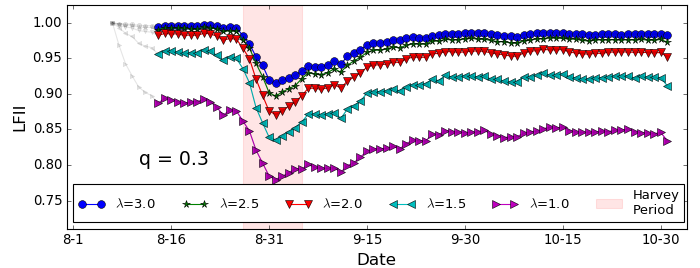}
\caption{\textbf{\textit{Link functionality irregularity index} change by date using different tolerance indices $\lambda$}. The congestion ratio $q$ is 0.3. The light-red shaded area indicates the period of Hurricane Harvey (August 26 to September 4)}
\label{fig:8-LFII to lambda}
\end{figure}

We examine the sensitivity of $LFII$ by changing $\lambda$ from 1.0 to 3.0 with the increments of 0.5, shown in Figure \ref{fig:8-LFII to lambda}. A larger $\lambda$ means more tolerance to link functionality fluctuation, and the curve of $LFII$ will be smoother. Beyond this range, however, it would be either too small to tolerate daily and weekly link functionality fluctuation or be too big to capture the irregular link when perturbation occurs. The $LFII$ curves using different $\lambda$ represent a standard resilience curve. When $\lambda$ increases, two boundaries are diverging away from the original state $DP_{oij}$, which will lead to more links assigned as regular links, resulting in a larger $LFII$ value. For example, when $\lambda$ is equal to 1.0, the original state of $LFII$ is 89\%;when $\lambda$ is equal to 3.0, the original state is almost 100\%. We also notice that before the Hurricane Harvey period, there is a small drop of $LFII$ from August 21 to August 23. Since the curve of $LFII$ is smooth, such drops (a decrease of 2.1\% when $\lambda$ is equal to 1.0) are significant and could be considered an early warning of the coming perturbation. 

\subsection{Resilience Quantification}

Based on the three network MOPs, we can examine and quantify the resilience of Harris County traffic network during urban flooding due to Hurricane Harvey using different values of congestion ratio $q$ and tolerance index $\lambda$. For network resilience analysis, we consider $t_d$ and $t_{ns}$ to be August 25 and October 2, respectively, and $t_r$ is the day with minimum MOP. Since we cannot avoid the fluctuation in the real case, we calculate the average MOP for the two-week period before $t_d$ as the initial steady-state MOP, and the one after $t_{ns}$ as the new steady-state MOP value. To compare all the MOPs, the original steady period value is normalized.

\begin{figure}[!ht]
\centering
\includegraphics[width=\linewidth]{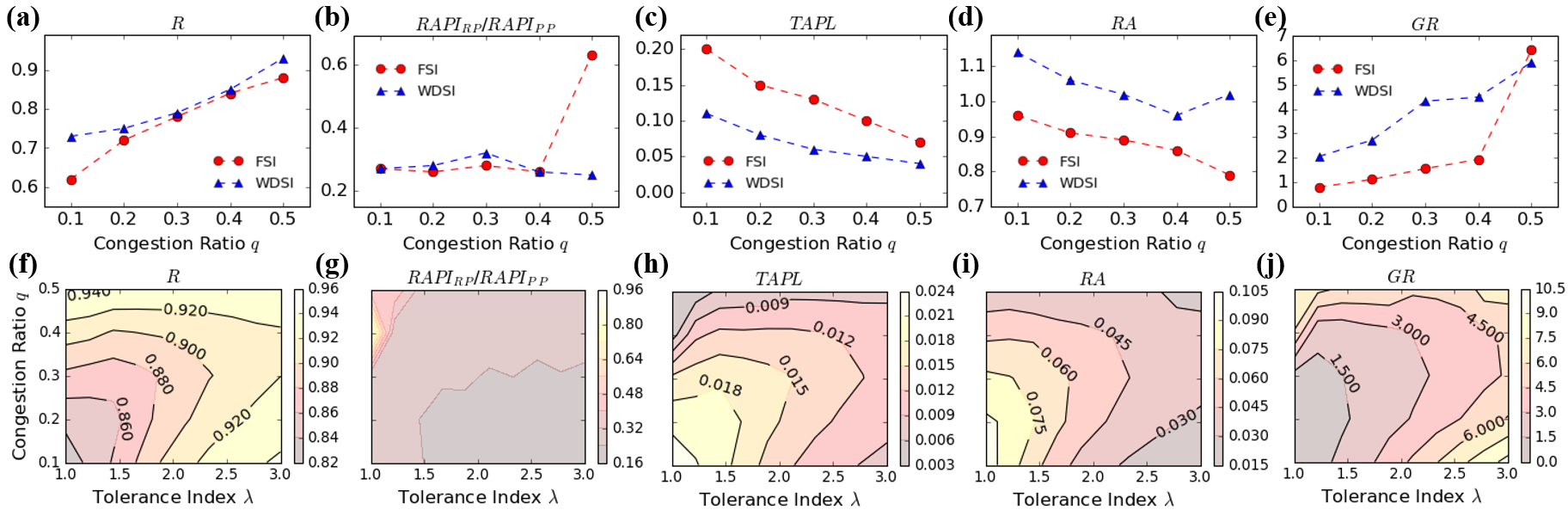}
\caption{\textbf{Relationship between three proposed MOPs and congestion ratio \textbf{\emph{q}} and tolerance index \textbf{\emph{$\lambda$}}.} \textbf{(a)-(e)} The $FSI$ and $WDSI$ change by the congestion ratio $q$. Red points represent $FSI$, blue points represent $WDSI$. \textbf{(f)-(j)} The $LFII$ change by the congestion ratio $q$ and the tolerance index $\lambda$.}
\label{fig:9-GR}
\end{figure}

Figure \ref{fig:9-GR} shows the variation of components of  the resilience metric based on the proposed MOPs (i.e., $FSI$, $WDSI$, and $LFII$) and their sensitivity to variations in $q$ or $\lambda$ values. Unlike \textit{FSI} and \textit{WDSI} with single parameter \textit{q}, the \textit{LFII} changes with both $q$ and $\lambda$, so they are plotted separately. Larger $q$ and $\lambda$ indicate less restriction in method capturing link dysfunction during analysis, so that network resilience would be better. The result shows that $R$ increases while $ATPL$ and $RA$ decrease with an increase in $q$ and $\lambda$, which is consistent with the relationship between metrics and resilience. The $RAPI_{RP}$/$RAPI_{PP}$ is more stable than other metrics for \textit{WDSI} and \textit{LFII}, and it increases significantly for \textit{FSI} when \textit{q} equals 0.5. The abnormal pattern above 0.5 results from the fact that the resilience curve of the proposed MOPs with large $q$ cannot reflect link dysfunction well in networks, as the difference between steady state and perturbed state is too small to be significant. 

In the calculation of network resilience, we combine all metrics to obtain $GR$. For each MOP, the value of $GR$ increases as $q$ or $\lambda$ increases. The sensitivity of $GR$ (calculated based on $FSI$ and $WDSI$ MOPs) to changes in $q$ or $\lambda$ diminishes when $q$ is larger than 0.5. Comparing the $GR$ among the three proposed MOPs, as shown in Figure \ref{fig:9-GR}(e) and (j), all show a positive relationship with $q$ and $\lambda$ in a reasonable range ($q \in [0,0.5]$ and $\lambda \in [1,3]$). $GR_{WDSI}$ is bigger than $GR_{FSI}$ suggesting that $WDSI$ shows more resilience than $FSI$, as the network topology contributes to network resilience in $WDSI$. 

\section{Discussion}

Many real-world networks have components with time-variant functionality, so the assessment of topological measures does not provide a full understanding of network resilience. This paper introduced three resilience measures of performance to capture important macroscopic, microscopic, and temporal characteristics of networks, considering both the topological structure of networks and the time-variant functionality of links. Illustrated by the case study of the Harris County traffic network before, during, and after Hurricane Harvey, the proposed measures of performance enable capture of resilience patterns in networks with time-variant functionality. $FSI$ and $WDSI$ are both macroscopic characteristics, but present different aspects of the network resilience and show different sensitivity to congestion ratio $q$ in steady state (Figure \ref{fig:6-different q}). $LFII$ focuses on the microscopic characteristic of network resilience by identifying all links with irregular behavior and calculating the proportion of regular links. 

Based on the evaluation of the three MOPs, the general network resilience ($GR$) is quantified for the period of external perturbation. The $GR$ obtained based on the $WDSI$ ($GR_{WDSI}$) MOP implies a greater resilience in the network compared with that from the $FSI$ ($GR_{FSI}$), since the network topology contributes to network resilience while $WDSI$ is used as the MOP. The $GR$ obtained based on the $LFII$ ($GR_{LFII}$) is more stable than the other two $GRs$. Each of the three MOPs provides a unique property for quantification of resilience based on network dynamics and topology. Specifically, the $FSI$ describes link functionality, while the $WDSI$ encapsulates both link functionality and network topology, which makes the $FSI$ more sensitive to the change of daily internal perturbations in link level. As for the $LFII$, the ability to capture individual link's dysfunctionality helps to identify vulnerable areas and links in order to allocate the resources and strategically optimize the network resilience. The proposed method is not limited to transportation/traffic networks. Other natural, physical, social, and engineered networks with time-variant functionality states and properties could also be examined and characterized using the proposed MOP indexes. Future research could further refine the theoretical approaches in the context of other networks and test them in the interdependent networks~\cite{Gao:2011:robustness}.

\bibliographystyle{unsrt}  
\bibliography{references}  %%% Remove comment to use the external .bib file (using bibtex).

\section*{Acknowledgements}

This material is based upon work supported by the National Science Foundation under award 1832662: CRISP 2.0 Type 2: Anatomy of Coupled Human-Infrastructure Systems Resilience to Urban Flooding: Integrated Assessment of Social, Institutional, and Physical Networks. This material is also based on the high-resolution traffic flow dataset provided by INRIX company. The authors are grateful of INRIX's support of this study. The authors would like to thank Chao Fan for providing suggestions to the paper. Their support is gratefully acknowledged.

\section*{Author contributions statement}

X.G., S.D., and A.M. conceived the research idea, X.G. conducted the experiment(s), X.G., S.D. and A.M. analysed the results. S.D., A.M., and J.G. advised on the experiments. A.M. (PI and supervisor) provided funding support. All authors reviewed and edited the manuscript.


%\begin{thebibliography}{1}

%\bibitem{kour2014real}
%George Kour and Raid Saabne.
%\newblock Real-time segmentation of on-line handwritten arabic script.
%\newblock In {\em Frontiers in Handwriting Recognition (ICFHR), 2014 14th
%  International Conference on}, pages 417--422. IEEE, 2014.

%\bibitem{kour2014fast}
%George Kour and Raid Saabne.
%\newblock Fast classification of handwritten on-line arabic characters.
%\newblock In {\em Soft Computing and Pattern Recognition (SoCPaR), 2014 6th
%  International Conference of}, pages 312--318. IEEE, 2014.

%\bibitem{hadash2018estimate}
%Guy Hadash, Einat Kermany, Boaz Carmeli, Ofer Lavi, George Kour, and Alon
%  Jacovi.
%\newblock Estimate and replace: A novel approach to integrating deep neural
%  networks with existing applications.
%\newblock {\em arXiv preprint arXiv:1804.09028}, 2018.

%\end{thebibliography}

\end{document}